\documentclass[showpacs,pra,twocolumn,amsmath]{revtex4-1} 

\usepackage{epsfig}
\usepackage{subfigure}
\usepackage{graphicx}
\usepackage{dcolumn}
\usepackage{stmaryrd}
\usepackage{mathrsfs}
\usepackage{pifont}
\usepackage{amsthm}
\usepackage{amssymb}
\usepackage{bm}
\usepackage{latexsym}
\usepackage[colorlinks=true,linkcolor=blue,citecolor=blue]{hyperref}
\usepackage{color}

\newcommand{\la}{\langle}
\newcommand{\ra}{\rangle}
\newcommand{\beq}{\begin{equation}}
\newcommand{\eeq}{\end{equation}}
\newcommand{\beqa}{\begin{eqnarray}}
\newcommand{\eeqa}{\end{eqnarray}}
\newcommand{\lam}{\lambda}
\newcommand{\ti}{\tilde}
\newcommand{\ga}{\gamma}

\newcommand{\da}{\dagger}

\newcommand{\si}{\sigma}

\newcommand{\om}{\omega}

\newcommand{\non}{\nonumber}

\def\jpa#1{{ J.\ Phys.\ A} {\bf#1}}
\def\pra#1{{ Phys.\ Rev. A\/} {\bf#1}}
\def\prb#1{{ Phys.\ Rev. B\/} {\bf#1}}
\def\prd#1{{ Phys.\ Rev. D\/} {\bf#1}}

\def\prl#1{{ Phys.\ Rev.\ Lett.} {\bf#1}}

\def\annph#1{{ Ann.\ Phys.} {\bf #1}}
\def\pla#1{{ Phys.\ Lett. A\/} {\bf#1}}
\def\rmp#1{{ Rev. \ Mod. \ Phys.} {\bf#1}}

\begin{document}
\title{Nonperturbative stochastic dynamics driven by strongly correlated colored noises}
\author{Jun Jing$^{1,2,3}$}
\thanks{J.J. and R.L. equally contributed to this work.}
\author{Rui Li$^{2}$}
\thanks{J.J. and R.L. equally  contributed to this work.}
\author{J. Q. You$^{2,4}$}
\thanks{Corresponding authors: J.Q.Y. (jqyou@csrc.ac.cn),  T.Y. (Ting.Yu@stevens.edu).}
\author{Ting Yu$^{1,2}$}
\thanks{Corresponding authors: J.Q.Y. (jqyou@csrc.ac.cn),  T.Y. (Ting.Yu@stevens.edu).}

\affiliation{$^1$Center for Controlled Quantum Systems and Department of Physics and Engineering Physics, Stevens Institute of Technology, Hoboken, New Jersey 07030, USA\\ $^2$Beijing Computational Science Research Center, Beijing 100084, China\\ $^3$Institute of Atomic and Molecule Physics, Jilin University, Changchun 130012, China\\ $^4$Synergetic Innovation Center of Quantum Information and Quantum Physics, University of Science and Technology of China, Hefei, Anhui 230026, China}

\date{\today}

\begin{abstract}
We propose a quantum model consisting of two remote qubits interacting with two correlated colored noises and establish an exact stochastic Schr\"odinger equation for this open quantum system. It is shown that the quantum dynamics of the qubit system is profoundly modulated by the mutual correlation between baths and the bath memory capability through dissipation and fluctuation. We report a new physical effect on generating inner-correlation and entanglement of two distant qubits arising from the strong bath-bath correlation.
\end{abstract}


\maketitle

\section{Introduction}

Nonequilibrium quantum open system involves understanding the quantum fluctuation and dissipation processes arising from the interaction between the open system and its surrounding environment \cite{CH}. Most researches on nonequilibrium dynamics have been focused on ambient noises from a single environment or several uncorrelated environments. However, in experimental environments a quantum system may be exposed to various kind of physical noises such as phase noise, amplitude noise, as well as classical noises. In many realistic situations, the participating noises are not totally statistically independent. Two distant atom-cavity systems connected by a fiber constitute an important example in the quantum network realization \cite{Mabuchietal}. Generally,  when the environment is formed by two or more mutually-interacting sub-environments, the correlations between the sub-environments will give rise to new physics on relaxation and decoherence across wide parameter regions interpolating totally uncorrelated individual environments and a single common environment. A deep understanding of the multi-correlation dynamical processes of open qubit systems holds enormous promise for both fundamental studies of non-Markovian quantum phenomena as well as practical applications to quantum information processing.

There are several approaches to solving quantum open systems coupled to an environment, including, e.g., the perturbative master equations \cite{Breuer1,Gurvitz}, the quantum trajectories \cite{quant2,quant3,quant4,quant5,quant6}, the dressed state method for simple open systems \cite{dress1}, Feynman-Vernon influence functional \cite{FV1,Legg,HPZ92}, and the projection operator approach that has been extensively used in statistical physics \cite{Nishi,Zwang}. As a new approach, a stochastic Schr\"odinger equation called non-Markovian quantum-state-diffusion (QSD) equation driven by a complex Gaussian noise was developed to describe an open quantum system that is coupled to bosonic environments, including individual bath \cite{PLA,Strunz99,Yupra99,JY} and common bath \cite{ZJBY}. However, when the acting environment is comprised of multi-correlated parts, each consisting of finite or infinite degrees of freedom, it is still a challenging task to establish a fully quantized theory to accurately describe the non-Markovian dynamics of the system, especially when the system-environment coupling is strong and when the acting environment cannot be approximated by a Markov bath \cite{www}. As far as we know, there are no fully rigorous investigations of either correlated non-Markovian environments or their effect on the entanglement of non-interacting qubit systems.

The purpose of this paper is to develop a nonperturbative approach to studying the nonequilibrium quantum dynamics of  an {\it exactly solvable} microscopic non-Markovian spin-dissipation model. We will consider an open system consisting of two non-interacting qubits coupled to an environment  formed by two baths that are correlated through interchanging quanta. Our research can be considered as a unifying treatment of the earlier research involving a single common environment (see, e.g., \cite{Braun,Paz,Mazzola}) and two uncorrelated environments \cite{Napoli,ZJBY,Simon,Ficek}).

Correlated dynamics inevitably involves multi-scale quantum dissipation and quantum backaction induced by the strong system-bath interaction. For a class of super-radiation state, we show that the bath-bath correlation will give rise to a crossover pattern exhibiting the competition between the bath memory parameter and the bath-bath coupling strength. Also, we show an exotic entanglement evolution of the two uncoupled qubits purely induced by correlated baths, including sudden collapse and revival of entanglement in the ultrastrong correlation regime of baths. Our nonperturbative quantum theory makes it feasible to study the non-Markovian dynamics of two separated qubits driven by strongly correlated colored noises, providing a microscopic tool to understand the new physical effects induced by the bath correlation.

The rest part of this paper is organized as follows. In Sec. \ref{model}, we apply the Bogoliubov transformation to obtain an exact QSD equation dealing with two central qubits coupled to two correlated baths. We calculate the inner-correlation dynamics of the two qubits in Sec. \ref{dynamics} to demonstrate the effect from the bath-bath correlation. In Sec. \ref{discuss}, we further discuss the application of the correlated baths on entangling two separated qubits. The conclusion is given in Sec. \ref{conclusion}. The details about the derivation of QSD equation can be  found in appendix \ref{supp}.

\section{The model and the exact QSD equation}\label{model}

Our proposed model of two central qubits dissipatively coupled to two correlated baths respectively is given by (setting $\hbar=1$)
\begin{eqnarray}\non
H_{\rm tot}&=&\frac{\omega_A}{2}\sigma^A_z+\frac{\omega_B}{2}\sigma^B_z+
\sum_k\omega_kd_k^\da d_k+\sum_k\omega_ke_k^\da e_k \\ \non
&+&\sum_k(\ti{g}_k\sigma^A_+d_k+\ti{g}_k^*\sigma^A_-d_k^\da)+
\sum_k(\ti{f}_k\sigma^B_+e_k+\ti{f}_k^*\sigma^B_-e_k^\da)\\ \label{Ht} &+&\sum_k\lambda_k(d_k^\da e_k+e_k^\da d_k),
\end{eqnarray}
where $\ti{g}_k$ ($\ti{f}_k$) is the coupling strength between qubit $A$ ($B$) and bath $d$ ($e$). For simplicity, we assume that (i) $\ti{f}_k=\kappa\ti{g}_k$, where $\kappa$ represents the anisotropy degree of the coupling parameter for the two qubits; (ii) both baths $d$ and $e$ are supposed to be at zero temperature; (iii) each $\lambda_k$ is the correlation strength between resonant modes of the wave vector $k$ in two baths and the correlations between off-resonant modes are neglected. It should be noted that this model can capture main features of a more generic correlated bath. For instance, it can describe the quantum-dynamical behaviors of two separated atoms in two quantum cavities (each in a cavity) mutually coupled via an optical fiber \cite{Fiber}. By using Bogoliubov transformation, $d_k=(a_k-b_k)/\sqrt{2}, \quad e_k=(a_k+b_k)/\sqrt{2}$, the original Hamiltonian can be converted to the following form:
\begin{eqnarray}\label{Htot}
H_{\rm tot}&=&H_S+H_I+H_R, \\ \non
H_S&=&\frac{\omega_A}{2}\sigma^A_z+\frac{\omega_B}{2}\sigma^B_z,
\\ \non H_I&=&\sum_kg_k^*[(\sigma^A_-+\kappa\sigma^B_-)a_k^\da+
(\sigma^A_--\kappa\sigma^B_-)b_k^\da]+{\rm h.c.},
\\ \non H_R&=&\sum_k\omega^a_ka_k^\da a_k+\sum_k\omega^b_kb_k^\da b_k,
\end{eqnarray}
where $\omega_k^a=\omega_k+\lambda_k$, $\omega_k^b=\omega_k-\lambda_k$, and $g_k=\ti{g}_k/\sqrt{2}$. Therefore, we have shown that the correlated baths can be mapped into two uncorrelated fictitious baths, each coupled to two qubits simultaneously. An important category of physical systems modeled by the above Hamiltonian includes, e.g., two capacitively coupled Cooper-pair boxes under charge fluctuations, which can be converted to two uncorrelated qubits experiencing correlated noises \cite{YHF} with an analog qubit-bath interaction Hamiltonian similar to Eq.~(\ref{Htot}). The coupling between the two physical baths $d$ and $e$ has been shifted into a modification to the frequencies in the new structured baths $a$ and $b$. It turns out that this transformation has paved a way of controlling the correlated dynamics through the parameter $\lambda_k$. Note that when $\lambda_k=0$, i.e., there is no direct interaction between the two baths, then the model simply reduces to the case with two separable baths \cite{Napoli}. When $|\lambda_k|=\omega_k$ (the resonant condition as well as the ultrastrong correlation regime), the model reduces to another important limiting case representing just one common bath \cite{ZJBY}.

It can be shown that the exact linear QSD equation, a time-local convolutionless equation, describing quantum trajectories by the Hamiltonian (\ref{Htot}) can be formally written as (see appendix \ref{supp})
\begin{equation}\label{LQSDO}
\partial_t\psi_t=(-iH_S+L_az_{at}
-L_a^\da\bar{O}_a+L_bz_{bt}-L_b^\da\bar{O}_b)\psi_t,
\end{equation}
where the two coupling operators are $L_a=\sigma^A_-+\kappa\sigma^B_-$, $L_b=\sigma^A_--\kappa\sigma^B_-$, and $\bar{O}_x\equiv\bar{O}_x(t,z_a,z_b)=\int_0^tds\alpha_x(t,s)O_x(t,s,z_a,z_b)$,  with $O_x$ explicitly given by
\begin{eqnarray}\non
O_x&=&f_{x1}(t,s)\sigma^A_-+f_{x2}(t,s)\sigma^B_-
+f_{x3}(t,s)\sigma^A_z\sigma^B_- \\ \non
&+&f_{x4}(t,s)\sigma^B_z\sigma^A_-+i\bigg[\int^t_0 ds' p_{xa}(t,s,s')z_{as'}
\\ \label{Oop} &+&\int^t_0 ds' p_{xb}(t,s,s')z_{bs'}\bigg]\sigma^A_-\sigma^B_-, \quad x=a,b.
\end{eqnarray}
The initial conditions for these $O$-operators are $O_x(s,s,z_a,z_b)=L_x$. Here $x=a,b$ denote the two baths (sources of environmental noise), $z_{xt}=-i\sum_kg_kz^*_{xk}e^{i\omega^x_kt}$ describes a time-dependent, complex Gaussian process that statistically satisfies $M[z_{xt}]=M[z_{xt}z_{xs}]=0$, and $M[z^*_{xt}z_{xs}]=\alpha_x(t,s)$, where $M[\cdot]$ stands for the ensemble average over the noise $z_{xt}$. It is important to note that $\alpha_x(t,s)$ is an {\it arbitrary correlation function} for bath $x$. Note that each $O$-operator explicitly contains the integrals over {\it noises from both baths} due to the fact that baths $a$ and $b$ are indirectly connected with each other through coupling to the system, a novel and important feature that does not emerge in the well-known cases with the local baths \cite{YEPRL04} or the common bath \cite{ZJBY}. Moreover, it is interesting to note that a set of partial differential equations for $f$'s (noise-free terms) and $p$'s (noise-integral terms) as well as their boundary conditions can be obtained by the QSD approach. To be more numerically efficient, the nonlinear QSD equation (see appendix \ref{supp} and Ref. \cite{Diosi}) for the normalized states,  $\ti{\psi}_t(z)=\frac{\psi_t(z)}{||\psi_t(z)||}$, is employed in the following numerical simulations.

\section{Non-Markovian nonequilibrium dynamics}\label{dynamics}

To show the crossover behavior of nonequilibrium dynamics from the Markov limit to the non-Markovian regime, the spectral density functions of the physical baths $d$ and $e$ are assumed to be identical and of a Lorentz form:  $S(\omega)=\frac{1}{2\pi}\frac{\ti{\Gamma}\gamma^2}{\gamma^2+\omega^2}$, where $\ti{\Gamma}$ is the coupling strength between each qubit system and its bath, $\gamma$ denotes the bandwidth of the bath, and $1/\gamma$ is proportional to the bath memory time. The correlation function is obtained via Fourier transform over $S(\om)$, 
\begin{equation}
\alpha(t,s)\equiv\sum_k|\ti{g}_k|^2e^{-i\omega_k(t-s)}=
\frac{\ti{\Gamma}\gamma}{2}e^{-\gamma|t-s|}
\end{equation}
When $\gamma$ approaches zero, the bath enters a very strong non-Markovian regime, and typically causes a non-Markovian system dynamics; when $\gamma$ becomes large or when time approaches infinity, $\alpha(t,s)\rightarrow\ti{\Gamma}\delta(t-s)$, recovering the well-known Markov limit. By definition, the correlation functions for the structured baths $a$ and $b$ are 
\begin{equation}\label{alphax0}
\alpha_x(t,s)=\sum_k|g_k|^2e^{-i\omega^x_k(t-s)}, \quad x=a,b. 
\end{equation} 
For simplicity, it is supposed that $\lambda_k=\lambda\omega_k$, where the dimensionless parameter $\lambda$ ($|\lambda|\leqslant1$) is used to measure the correlation strength between the physical baths. It is easy to check that when $|\lambda|<1$, 
\begin{equation}\label{alphax}
\alpha_x(t,s)=\frac{\Gamma\gamma}{2}e^{-\gamma_x|t-s|},
\end{equation} 
where $\Gamma=\ti{\Gamma}/2$, $\gamma_a=\gamma(1+\lambda)$, and $\gamma_b=\gamma(1-\lambda)$; when $|\lambda|=1$, the correlation function of the {\em effective} common bath is $\frac{\Gamma\gamma}{2}e^{-2\gamma|t-s|}$. Here we do not consider the case with $|\lam|>1$. Otherwise the system may become unphysical and lose its positivity.

With the nonlinear QSD equation (\ref{nonli}), the correlated Gaussian noises $z_{xt}$ and the time-evolution function for $\bar{O}_x$, we can efficiently simulate the exact dynamics of the central qubit system by solving the QSD equation: $\rho_t=M[|\psi_t\ra\la\psi_t|]$. Below we discuss the nonequilibrium quantum dynamics by examining the inner-correlation $C_{xx}\equiv\la\sigma_x^A\sigma_x^B\ra$ and entanglement $C(t)$ measured by concurrence \cite{con} of the two-qubit system. The parameters are chosen as $\kappa=1$, and $\Gamma=\omega_A=\omega_B=\omega$,  (corresponding to a strong coupling regime between qubits and baths), and the results are obtained by average over $1000$ quantum trajectories, which are sufficient for the QSD equation to attain convergence for our model.

\begin{figure}[htbp]
\centering
\includegraphics[width=3.2in]{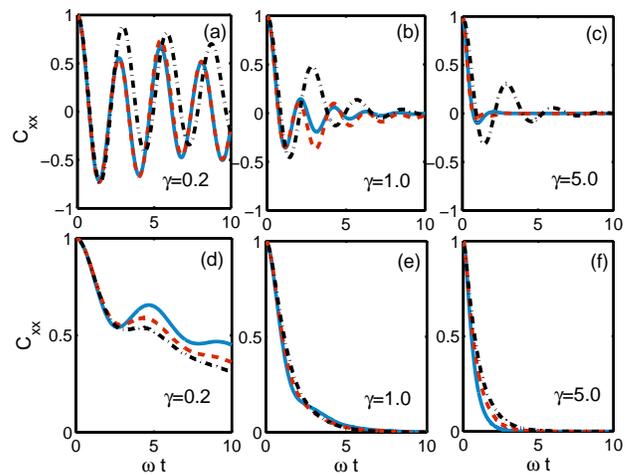}
\caption{(Color online) Inner-correlation dynamics with different values of $\gamma$ and positive bath-bath correlation strength (blue solid line for $\lambda=0.2$, red dashed line for $\lambda=0.6$, and black dot-dashed line for $\lambda=1.0$). In (a), (b) and (c), $\psi_0=(1/\sqrt{2})(|11\rangle+|00\rangle)$; in (d), (e) and (f), $\psi_0=(1/\sqrt{2})(|10\rangle+|01\rangle)$.} \label{plam}
\end{figure}

In Fig.~\ref{plam}, we observe the behavior of the inner-correlation between qubit $A$ and qubit $B$ along the $x$-direction, $C_{xx}$, describing a collective property relevant to the system coherence, with different initial entangled states, environment memory times and positive correlation strengths ($0<\lam\leqslant1$) between the baths $d$ and $e$. The system is initially prepared in the two-photon entangled state $(1/\sqrt{2})(|11\rangle+|00\rangle)$ and the single-photon entangled state $(1/\sqrt{2})(|10\rangle+|01\rangle)$, respectively. In a strong non-Markovian regime with $\gamma=0.2$, the system inner-correlation shows oscillations stronger in the two-photon entangled state than in the single-photon entangled state [comparing Figs.~\ref{plam}(a) and \ref{plam}(d)]. For the two-photon entangled state, a larger $\lambda$ gives rise to a much stronger oscillations. But for the single-photon entangled state, the dynamical pattern induced by increasing $\lambda$ is reversed. In a moderate range $\gamma=1.0$, for $\psi_0=(1/\sqrt{2})(|11\rangle+|00\rangle)$ [see Fig.~\ref{plam}(b)], only small bumps appear in the inner-correlation decay process when $\lambda=0.6$. The resonant condition $\lambda=1$ can extend the survival time of the inner-correlation. While for the state $(1/\sqrt{2})(|10\rangle+|01\rangle)$ [see Fig.~\ref{plam}(e)], the dynamics is almost $\lambda$-independent. When $\gamma=5.0$, both baths are effectively in the Markov regime, i.e., the inner-correlation quickly decays into zero monotonously, irrespective of the initial states [see Figs.~\ref{plam}(c) and \ref{plam}(f)]. However, we notice that, as a rather counterintuitive result shown in Fig.~\ref{plam}(f), the survival time of the inner-correlation in the resonant condition ($\lambda=1$) is longer than those in the off-resonant cases. It is because the two qubits experience an effective common bath at $\lambda=1$, and the single-photon state $(1/\sqrt{2})(|10\rangle+|01\rangle)$ serves as a super-radiant state due to the coupling terms in Eq.~(\ref{Htot}) \cite{Maniscalco}. Therefore,  the effect of bath-bath correlations on system dynamics is {\it state-dependent}.

To get a better picture of the nonequilibrium processes of the qubit system, it is useful to consider the Markov limit of  this correlated-baths model with the bath correlation functions $\alpha_x(t,s)=\Gamma_x\delta(t-s)$, where $\Gamma_a=\Gamma/(1+\lambda)$ and $\Gamma_b=\Gamma/(1-\lambda)$. Consequently, for $\lambda<1$, the Lindblad master equation should be written as $
\partial_t\rho_t=[-iH_S,\rho_t]+\sum_{x=a,b}\Gamma_x\mathcal{D}(L_x)$,  where $\mathcal{D}(L_x)\equiv L_x\rho_tL_x^\da-
\frac{1}{2}L_x^\da L_x\rho_t-\frac{1}{2}\rho_tL_x^\da L_x$, and for $\lambda=1$, $\partial_t\rho_t=[-iH_S,\rho_t]+\frac{\Gamma}{2}\mathcal{D}(L_a)$.  Considering the special initial state $(1/\sqrt{2})(|10\rangle+|01\rangle)$ (the eigenstate of $L_b$), one can see that the damping rate of the system correlation is $2\Gamma_a=\frac{2}{1+\lambda}\Gamma$, which increases with decreasing $\lambda$ and is larger than the damping rate $\Gamma$ for the $\lam=1$ case. Therefore, in this situation, the $\lam$-influence on inner-correlation in Markov limit [see Fig.~\ref{plam}(f)] is opposite to that  demonstrated in a highly non-Markovian case [see Fig.~\ref{plam}(d)]. This observation is helpful to understand the insensitivity of the dynamics to $\lambda$ in the moderate non-Markovian regime [see Fig.~\ref{plam}(e)], where the crossover pattern reflects the tradeoff between the two competing elements: $\lambda$ and $\gamma$.

\section{Entangling distant qubits via bath correlations}\label{discuss}

\begin{figure}[htbp]
\centering
\includegraphics[width=3.2in]{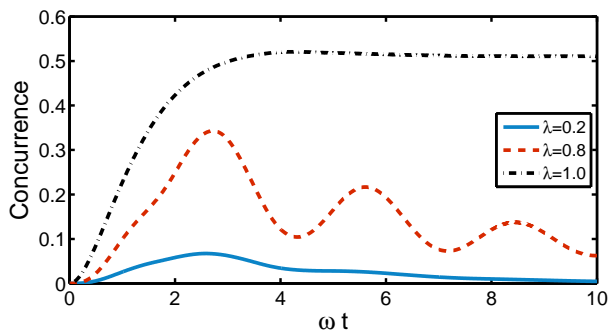}
\caption{ (Color online) Concurrence of distant qubits under the modulation of correlated bath with different correlation strength $\lambda$. Here $\psi_0=|10\rangle$ and $\ga=1.0$.} \label{nlam}
\end{figure}

A prompt application of this study is to dynamically entangle two remote qubits. To focus exclusively on the effects on the entanglement generation of the qubits as it arises from the bath correlation, in Fig.~\ref{nlam}, we keep $\ga$ (a measurement of  the memory capability for both baths) unchanged, and tune the bath-bath correlation strength $\lambda$ to display the entangling process of the two-qubit system that is initially prepared in the separable state $\psi_0=|10\rangle$. Apart from the well-known ultrastrong correlation, i.e., the effective common bath case with $\lam=1$, we show that the remote qubits can also be temporarily entangled purely via bath correlation in a wide parameter range. More specifically, for a weak bath correlation $\lam=0.2$ (the blue solid curve in Fig.~\ref{nlam}), the generated concurrence of the two-qubit system is present albeit small and has weak oscillations. When $\lam$ increases (e.g., $\lam=0.8$), the generated concurrence becomes large and exhibits appreciable oscillations (see the red dashed curve in Fig.~\ref{nlam}). Moreover, it can be seen that the concurrence at $\lam=0.2$ and $0.8$ tends to zero when $t\rightarrow\infty$. This is because the reduced density operator of the system approaches $\rho_\infty=|00\ra\la00|$ in the long-time limit for any $\lam<1.0$. However when $\lam=1.0$, which corresponds to the two qubits coupled to an effective common bath, the generated concurrence increases faster and then maintains at a given value of $0.5$. The reason for the observed results is that the reduced density operator of the system approaches $\rho_\infty=\frac{1}{2}|\phi\ra\la\phi|+\frac{1}{2}|00\ra\la00|$ in the long-time limit for such a common-bath case, where $|\phi\ra=(1/\sqrt{2})(|10\ra-|01\ra)$ is an eigenstate of $L_a=\si_-^A+\si_-^B$, which is the only effective Lindblad operator in the ultrastrong correlation case.

\begin{figure}[htbp]
\centering
\includegraphics[width=3.2in]{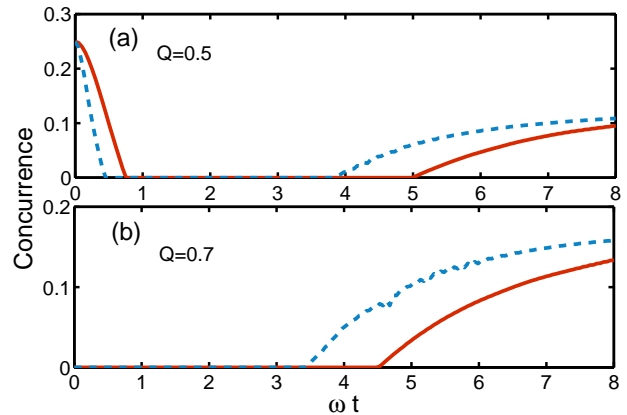}
\caption{(Color online) Temporal entanglement behaviors with different values of parameter $Q$ for generalized Werner-state in the ultrastrong correlated bath. The red solid line corresponds to $\gamma=1.0$ and the blue dashed line to $\gamma=5.0$.} \label{gere}
\end{figure}

Next, we focus on the exotic behaviors of the quantum entanglement induced by the {\it ultrastrong} bath correlation with $\lam=1.0$ which has not been nonperturbatively studied before. We choose a generalized Werner-state as the initial state: $\rho_0=\frac{Q}{4}I_4+(1-Q)|\psi\rangle\langle\psi|$, where $0\leqslant Q<1$, $I_4$ is the identity matrix in the Hilbert space of the two-qubit system, and $|\psi\rangle=(1/\sqrt{2})(|10\rangle+|01\rangle)$. It is known that the initial concurrence is $C(0)=\max\{0,1-\frac{3}{2}Q\}$. Apparently, when $Q <\frac{2}{3}$, the two-qubit system is initially entangled. In this case, the system suffers a fast entanglement decay in a short time [see Fig.~\ref{gere}(a)]. Moreover, the entanglement lifetime becomes significantly shortened when increasing the bath memory parameter $\ga$, which means that {\it the destructive effect on system entanglement from the Markov bath is more serious than a comparatively non-Markovian bath}. After a period of time, the entanglement suddenly revives due to the accumulation of correlation between the two baths. Also, as shown in Fig.~\ref{gere}(a), a larger $\ga$ yields an earlier revival of entanglement. This correlation-induced behavior is in sharp contrast to the usual observation that a non-Markovian bath is helpful in enhancing the coherence refocusing of system compared with a Markov bath. When $Q\geqslant\frac{2}{3}$, then the two qubits evolve from a separable mixed state. In this case, only the generation of the quantum entanglement occurs [see Fig.~\ref{gere}(b)]. In the long-time limit, the reduced density operator of the system approaches $\rho_\infty=\frac{Q}{4}|\phi\ra\la\phi|+(1-\frac{Q}{4})|00\ra\la00|$, so $C(\infty)=Q/4$. Therefore, all of the curves in Fig.~\ref{gere} will reach this value, implying that the steady state is irrelevant to $\ga$. It turns out that the generated entanglement degree must be larger than its original value when $Q>\frac{4}{7}$. This provides a method to increase the quantum entanglement of two distant qubits via bath correlation rather than the qubit-qubit coupling or a common bath.

\section{Conclusion}\label{conclusion}

We have presented here an investigation of the exact dynamical equation for two uncoupled qubits driven by strongly correlated noises. We have observed new physical phenomena arising from the bath-bath correlation. Particularly, we have observed the crossover dynamics dictated by two competitive parameters of the bath memory capability $1/\ga$ and the bath-bath correlation strength $\lam$. We have shown that the entanglement evolution of two qubits can be affected in several ways, depending on the initial states, environment memory time, and the bath-bath correlation. In particular, it is shown that the bath-bath correlation in a realistic context such as distant atoms in a quantum network can play a mediator role in generating entanglement of two remote qubits. Our research has shed new light on the fundamental studies in the nonequilibrium open quantum dynamics.  In addition, the treatment and results presented here are expected to be applicable to quantum devices involving multi-correlated baths.

\begin{acknowledgements}
We acknowledge grant support from the NSF No.~PHY-0925174, the NSAF No. U1330201, the NBRPC No.~2014CB921401, the NSFC Nos.~91421102 and 11175110, and Science and Technology Development Program of Jilin Province of China (20150519021JH).
\end{acknowledgements}

\appendix

\section{Exact QSD equation for two-qubit system driven by correlated baths}\label{supp}

The purpose of this appendix is to provide the derivation of the linear (\ref{LQSDO}) and non-linear quantum-state-diffusion (QSD) equations \cite{quant4} and the details of the explicit constructions of the O-operators for the model of two distant qubits coupled to two correlated baths presented in the main text.

Via Bogoliubov transformation, the model where an open quantum system interacts with two correlated baths (\ref{Ht}) is equivalent to that where the system is coupled with two uncorrelated effective baths simultaneously (\ref{Htot}). When the correlation strength parameter $\lam\in(-1, 1)$, we have two independent correlation functions for the baths $a$ and $b$ as given in Eq.~(\ref{alphax0}).

In the framework of non-Markovian quantum trajectory method \cite{PLA,Diosi,Strunz99}, it is shown that in the linear QSD equation (\ref{LQSDO}), two O-operators are introduced by:
\begin{equation*}
O_x(t,s,z_a,z_b)\psi_t\equiv\frac{\delta}{\delta z_{xs}}\psi_t, \quad
z_{xs}=-i\sum_kg_kz_{xk}^*e^{i\omega_k^xs},
\end{equation*}
with the initial conditions:
\begin{eqnarray}\non
O_a(s,s,z_a,z_b)&\equiv&L_a=\sigma^A_-+\kappa\sigma^B_-, \\ \label{init}
O_b(s,s,z_a,z_b)&\equiv&L_b=\sigma^A_--\kappa\sigma^B_-.
\end{eqnarray}
And the nonlinear QSD equation is
\begin{eqnarray}\label{nonli}
&& \frac{\partial}{\partial t}\ti{\psi}_t=\big\{-iH_S+\sum_{x=a,b}\big[\Delta_t(L_x)\ti{z}_{xt} \\ \non
&&-\Delta_t(L_x^\da)\bar{O}_x(t,\ti{z}_a,\ti{z}_b)
+\langle\Delta_t(L_x^\da)\bar{O}_x(t,\ti{z}_a,\ti{z}_b)\rangle_t\big]\big\}
\ti{\psi}_t,
\end{eqnarray}
where
\begin{eqnarray*}
\bar{O}_x(t,z_a,z_b)&\equiv&\int_0^tds\alpha_x(t,s)O_x(t,s,z_a,z_b), \\
\Delta_t(A)&\equiv&A-\langle A\rangle_t, \quad
\langle A\rangle_t\equiv\langle\ti{\psi}_t|A|\ti{\psi}_t\rangle \\
\ti{z}_{xt}&=&z_{xt}+\int_0^t\alpha_x(t,s)\langle L_x^\da\rangle_sds.
\end{eqnarray*}
By the consistency conditions: $\frac{\partial}{\partial t}
\frac{\delta}{\delta z_{xs}}\psi_t=\frac{\delta}{\delta z_{xs}}
\frac{\partial}{\partial t}\psi_t$, we have
\begin{eqnarray} \non
\frac{\partial}{\partial t}O_x
&=&[-iH_S+\sum_{y=a,b}(L_yz_{yt}-L_y^\da\bar{O}_y),
O_x]\\ \label{CC} &-&\sum_{y=a,b}L_y^\da\frac{\delta\bar{O}_y}{\delta z_{xs}}
\end{eqnarray}
Equation (\ref{CC}) yields the exact O-operators that can be explicitly constructed as Eq.~(\ref{Oop}). According to Eq.~(\ref{init}), the initial conditions for the coefficient functions $f$'s are expressed by
\begin{equation*}
f_{a1}(t,t)=1, f_{a2}(t,t)=\kappa, \quad f_{b1}(t,t)=1, f_{b2}(t,t)=-\kappa.
\end{equation*}
Defining $F_{xj}(t)\equiv\int_0^t\alpha_x(t,s)f_{xj}(t,s)ds$, and
$P_{xy}(t,s')\equiv\int_0^t\alpha_x(t,s)p_{xy}(t,s,s')ds$, $x,y=a,b$, it follows that
\begin{eqnarray}\label{Oxbar}
&& \bar{O}_x(t,z_a,z_b)=F_{x1}(t)\sigma^A_-+F_{x2}(t)\sigma^B_- \\ \non
&&+ F_{x3}(t)\sigma^A_z\sigma^B_-+F_{x4}(t)\sigma^B_z\sigma^A_- \\
\non &&+ i\left[\int^t_0 ds' P_{xa}(t,s')z_{as'}
+\int^t_0 ds' P_{xb}(t,s')z_{bs'}\right]\sigma^A_-\sigma^B_-.
\end{eqnarray}
Inserting these expressions of O-operators into Eq.~(\ref{CC}), we can easily obtain the partial differential equations for $f_{xj}$'s, with $x=a,b$, $j=1,2,3,4$; $p_{xy}$'s, with $x,y=a,b$, and the boundary conditions:
\begin{eqnarray*}
p_{xa}(t,s,t)&=&-2if_{x3}(t,s)-2i\kappa f_{x4}(t,s), \\
p_{xb}(t,s,t)&=&-2if_{x3}(t,s)+2i\kappa f_{x4}(t,s).
\end{eqnarray*}

Note that the equations for $f_{xj}$'s and $p_{xy}$'s are valid for bosonic baths with arbitrary correlation functions or spectral density functions. With Eqs.~(\ref{Oop}), (\ref{alphax}), (\ref{CC}), and (\ref{Oxbar}), we can find a closed set of the ordinary differential equations for the coefficients of $\bar{O}_x$ for the sake of numerical simulation over Eq.~(\ref{nonli}).

\end{document}